\documentclass[twocolumn,aps,showpacs,showkeys,prb,superscriptaddress,reprint]{revtex4-1}
\usepackage{epsfig}
\usepackage{graphicx}
\usepackage{amsmath}
\usepackage{amssymb}
\usepackage{amsfonts}
\usepackage{tabularx}
\usepackage{color}
\usepackage{dcolumn}
\usepackage{bm}

\begin{document}

\title{Convergence of the Bloch-waves method}

\author{J\'{a}n Rusz}
\affiliation{Department of Physics, Uppsala University, Box 530, S-751~21 Uppsala, Sweden}
\affiliation{Institute of Physics, Academy of Sciences of the Czech Republic, Na Slovance 2, CZ-182 21 Prague, Czech Republic}

\author{Shunsuke Muto}
\author{Kazuyoshi Tatsumi}
\affiliation{Department of Materials, Physics, and Energy Engineering, Nagoya University, Chikusa, Nagoya 464-8603, Japan}

\begin{abstract}
We discuss in detail the Bloch waves method for calculation of energy and orientation dependent scattering cross-section for inelastic scattering of electrons on crystals. Convergence properties are investigated and a new algorithm with superior timing and accuracy is described. The new method is applied to calculations of intensity of weakly excited spots, maps of magnetic signal, and tilt series from zone axis orientation towards three-beam orientation.
\end{abstract}


\maketitle

\section{Introduction}

Dynamical diffraction theory describes the multiple elastic scattering of electrons through a crystal. The theory has been formulated almost a century ago, originally for photons by Bragg and von Laue. Later it was extended also to electrons on the basis of wave-particle duality, as proven by the famous experiments of Davidson and Germer, and Thomson and Reid. The importance of inelastic electron scattering and the wealth of information contained in this process was for the first time observed by Kikuchi\cite{kikuchi}, who observed a complicated pattern of lines going beyond the diffraction patterns expected for elastic scattering of light or electrons. The diffraction theory that included quantitative description of Kikuchi patterns was given by Kainuma following qualitative theories of Shinohara and von Laue\cite{kainuma}. Various sources of energy loss, such as excitations of phonons or valence electrons to conduction band or core electrons to conduction band were discussed by Okamoto et al.\cite{okamoto}

The inelastic electron scattering is typically described as a three-step process: 1) an elastic scattering of the probe electron from entrance surface to a selected atom in the sample, 2) an inelastic event exchanging the momentum and energy between a probe and sample electron, and 3) an elastic propagation of the scattered probe electron towards the exit sample surface. This needs to be summed over all possible inelastic scattering centers. Such description holds, when there is only one inelastic scattering event for each observed probe electron. If its mean free path is much longer than the thickness of the sample, then it is a good approximation. Otherwise one needs to consider multiple inelastic scattering\cite{egerton}.

The inelastic event is described by mixed dynamical form-factor (MDFF) introduced by Kohl \cite{kohlrose}. For calculations of MDFF, there are various levels of sophistication, ranging from an isotropic dipole approximation $\mathbf{q} \cdot \mathbf{q'}$, through a parametrized atomic multiplets description \cite{lionel}, configuration interaction \cite{kazu}, to a density functional theory evaluation \cite{nelhiebel, prbtheory}. Recently a DFT-based dipole model for MDFFs has been published in \cite{opmaps} using local electronic structure properties to set the coefficients in the dipole approximation, thereby providing a realistic DFT-based MDFF model with efficiency equal to a simple dipole approximation (at the cost of losing the fine structure in energy dependence).

The elastic scattering can be simulated by one of the two major methods widely used today, the multislice method\cite{cowley,ishizuka} and Bloch-waves (BW) method\cite{inokuti,metherell}. The latter one is typically more efficient for periodic structures without defects, while the multislice method has an advantage when dealing with large non-periodic structures or structures with defects. In this article we will study the convergence properties of the BW method in electron energy-loss near edge structure (ELNES) calculations and describe an efficient algorithm for BW summation.

\section{Bloch-waves theory of ELNES}

In Ref.~\onlinecite{prbtheory} we have described our theoretical approach to simulate general orientation-sensitive ELNES experiment from first principles. For the sake of completeness, we present here a brief summary of the key equations.

The first step is a solution of the secular equation for a fast electron of known energy moving in and out of the crystal. Expanding the solutions - Bloch waves - into plane waves (indexed by reciprocal lattice vectors $\mathbf{g},\mathbf{h}$) the secular equation has the following form:
\begin{equation}
  \sum_\mathbf{g} \bigg[ \left( K^2 - (\mathbf{k}^{(j)}+\mathbf{g})^2 \right) + \sum_{\mathbf{h} \ne 0} U_\mathbf{h} C_{\mathbf{g}-\mathbf{h}}^{(j)} \bigg] e^{i(\mathbf{k}^{(j)}+\mathbf{g}) \cdot \mathbf{r}}=0
\end{equation}
where $K^2 = U_0 + 2meE/\hbar^2$, $m$ and $e$ are, respectively, the electron mass and charge, $U_\mathbf{g}=2meV_\mathbf{g}/\hbar^2$ where $V_\mathbf{g}$ are the Fourier components of the crystal potential, and $\mathbf{k}^{(j)} = \mathbf{k} + \gamma^{(j)}\mathbf{n}$ are Bloch vectors related to the beam direction $\mathbf{k}$ and sample surface normal $\mathbf{n}$. 

Solution of the secular equation is a set of Bloch waves indexed by $j$, given by so called Bloch coefficients $C_\mathbf{g}^{(j)}$ and elongations of the wave vector $\gamma^{(j)}$.

Evaluation of the double differential scattering cross-section involves calculation of the following sum [see, e.g., Ref.~\onlinecite{prbtheory}, Eq.~(24)]
\begin{eqnarray} \label{eq:dscsfin}
  \frac{\partial^2 \sigma}{\partial\Omega \partial E} & = & \sum_{\mathbf{ghg}'\mathbf{h}'}
    \frac{1}{N_\mathbf{u}} \sum_\mathbf{u}
                                           \frac{S_\mathbf{u}(\mathbf{q},\mathbf{q'},E)}{q^2 q'^2} e^{i(\mathbf{q}-\mathbf{q'})\cdot\mathbf{u}}
                      \nonumber \\  & \times &
    \sum_{jlj'l'} Y_{\mathbf{ghg}'\mathbf{h}'}^{jlj'l'} T_{jlj'l'}(t)
\end{eqnarray}
where
\begin{eqnarray}
  Y_{\mathbf{ghg}'\mathbf{h}'}^{jlj'l'} & = &
   C_{\mathbf{0}}^{(j)\star} C_{\mathbf{g}}^{(j)}
   D_{\mathbf{0}}^{(l)} D_{\mathbf{h}}^{(l)\star}
                       \\  & \times &
   C_{\mathbf{0}}^{(j')} C_{\mathbf{g'}}^{(j')\star}
   D_{\mathbf{0}}^{(l')\star} D_{\mathbf{h'}}^{(l')} .
                       \nonumber
\end{eqnarray}
Here $C_\mathbf{g}^{(j)}$ and $D_\mathbf{h}^{(l)}$ are the Bloch coefficients for the incoming and outgoing Bloch fields. The quantity $\frac{S_\mathbf{u}(\mathbf{q},\mathbf{q'},E)}{q^2 q'^2}$ is the mixed-dynamic form-factor (MDFF) divided by squares of momentum transfer vectors $\mathbf{q},\mathbf{q'}$ (Coulomb potential factors). $T_{jlj'l'}(t)$ is a thickness function, which depends on Bloch wave indices and experimental geometry. $N_\mathbf{u}$ is a number of atoms in the unit cell, where $\mathbf{u}$ is a base vector. Momentum transfer vectors actually depend on several indices, but to simplify the notation we will not write this dependence explicity. For evaluation of MDFFs we typically use an approximation
\begin{equation}
  \mathbf{q} = \mathbf{k}_\mathrm{out}^{(l)} - \mathbf{k}_\mathrm{in}^{(j)} + \mathbf{h} - \mathbf{g} \approx \mathbf{k}_\mathrm{out} - \mathbf{k}_\mathrm{in} + \mathbf{h} - \mathbf{g}
\end{equation}
which neglects elongations of the wave vectors $\gamma^{(j,l)}$.

\section{Summation algorithms}

Here we will discuss two known algorithms used in Bloch wave calculations and propose new algorithms for improved convergence and efficiency of summation.

\subsection{Manual selection of beams\label{sec:adhoc}}

The simplest one is based on a choice of beams ``by hand'' let's say on a base of visible spots in diffraction pattern, or all beams from a set of integer indices below certain cut-off, or beams on a systematic row, etc. For example, in \cite{prbtheory,cobalt} we used a systematic row approximation both for secular equation and summation, i.e., we picked only a set of $\sim 10$ beams along the systematic row of reflections. Pragmatic hand-selection of beams is widespread in literature, see for example Refs.\cite{allen,rossouw,levine,tatsumi,kirkland}.

\subsection{Excitation error and extinction distance based selection of beams\label{sec:wgbs}}

In our previous work \cite{prbtheory} we suggested to choose beams on the base of their excitation error $s_\mathbf{g}$ and extinction distances $\xi_\mathbf{g}$. Their product forms a dimensionless variable $w_\mathbf{g}$, for which we set a cut-off criterion. This method picks beams that follow the Ewald sphere and provides automatically a more economic and accurate description. A variant of this method uses two different cut-offs for the $w_\mathbf{g}$ - one for the secular equation (hundreds or thousands of beams) and one for the summation (10--15 beams). We can safely take a rather large set of beams in the first step -- the solution of the secular equation -- and then for the summation we pick only a subset of the beams and Bloch waves. The selection of the subset of Bloch waves is based on their norm within the subspace of beams selected for summation\cite{prbtheory}. This has been used in our more recent publications, e.g., \cite{snr,cobalt,emcd2nm,hansprl}.

Both this approach and the manual selection of beams identify a set of beams and Bloch waves and then they count all the cross-terms. I.e., if we have $N_\mathbf{g}$ beams and $N_j$ Bloch waves, the summation runs over $N_\mathbf{g}^4 N_j^4$ terms, which can be a huge number already for only 10 $\mathbf{g}$-vectors and 10 Bloch waves. A detailed inspection of these terms shows, that majority of them are actually negligible. That means, that we sum a lot of tiny terms, which is not only inefficient but also contributes to the propagation of machine rounding off errors.

\begin{figure}[htb]
  \includegraphics[width=8.5cm]{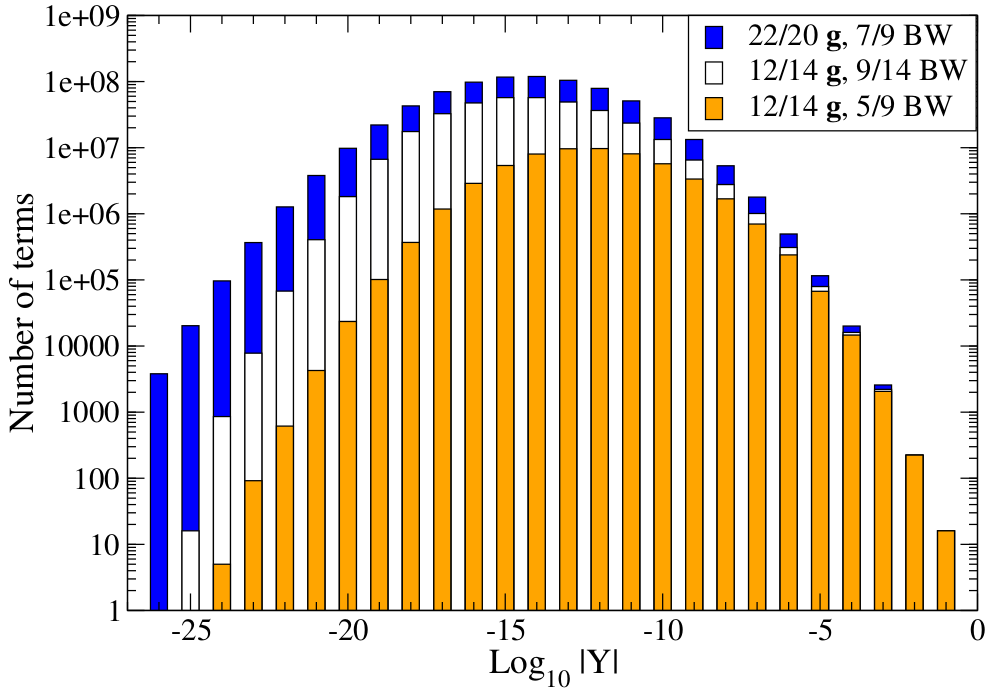}
  \caption{Histogram of the distribution of $Y_\mathbf{ghg'h'}^{jlj'l'}$ terms magnitudes as a function of number of $\mathbf{g}$-vectors and Bloch waves (BW). \label{fig:hist}}
\end{figure}

To illustrate the finding, we have set-up test calculations in a 3-beam orientation with $\mathbf{G}=(200)$ and detector orientation $(\frac{G}{2},\frac{G}{2})$. By setting the maximum $w_\mathbf{g}=10^5$ for the secular equation, the program identified a list of 628 and 627 $\mathbf{g}$-vectors for incoming and outgoing beams, respectively. For the summation we set the maximum $w_\mathbf{g}=500$, which filtered the list down to 12 and 14 $\mathbf{g}$-vectors. These contained $[000]$, $\pm[200]$, $\pm[110]$, $\pm[1\bar{1}0]$, $[310]$, $[\bar{3}10]$, $\pm[400]$ and $[020]$. The outgoing beam included on top of that $[0\bar{2}0]$ and $[\bar{5}10]$. With the criterion of minimal norm of the Bloch wave in the subspace of identified beams $0.01$ we got 5 and 9 Bloch waves, while for criterion $0.005$ we got 9 and 14 Bloch waves for incoming and outgoing beam, respectively. In total that makes above 57 million vs 354 million terms, per energy and thickness. The computing time of one thickness profile on a single 2.0~GHz Intel Pentium 4 Xeon CPU was 45s vs 285s, respectively. 

Instead of increasing number of Bloch waves, we also tested the effect of enlarging the set of $\mathbf{g}$-vectors in summation by setting $w_\mathbf{g}=800$. The number of selected $\mathbf{g}$-vecotrs grows to 22 and 20, and some of the $hkl$ beams have nonzero $l$ (higher order Laue zones). Keeping minimal norm for Bloch waves $0.01$ we got 7 and 9 Bloch waves. In total it gives 768 milions of terms, which were summed in 3886s.

Figure~\ref{fig:hist} shows histograms of distributions of the sizes of all these terms. It turns out that if we want to assure that all of the, let's say, 1000 dominant terms (for our fixed selection of beams) are included in the summation, we are also including into the sum many millions of terms with much smaller magnitudes, majority of them having negligible influence on results. As we will show below, 1000 dominant terms might not always be enough in terms of convergence. Then one can easily conclude that requiring a summation over, e.g., $10^5$ dominant terms would require such a number of beams and Bloch waves that we would end up summing billions of terms, greatly wasting computational resources.

\subsection{Automatic selection of dominant terms\label{sec:ats}}

\begin{figure}
  \includegraphics[width=8.5cm]{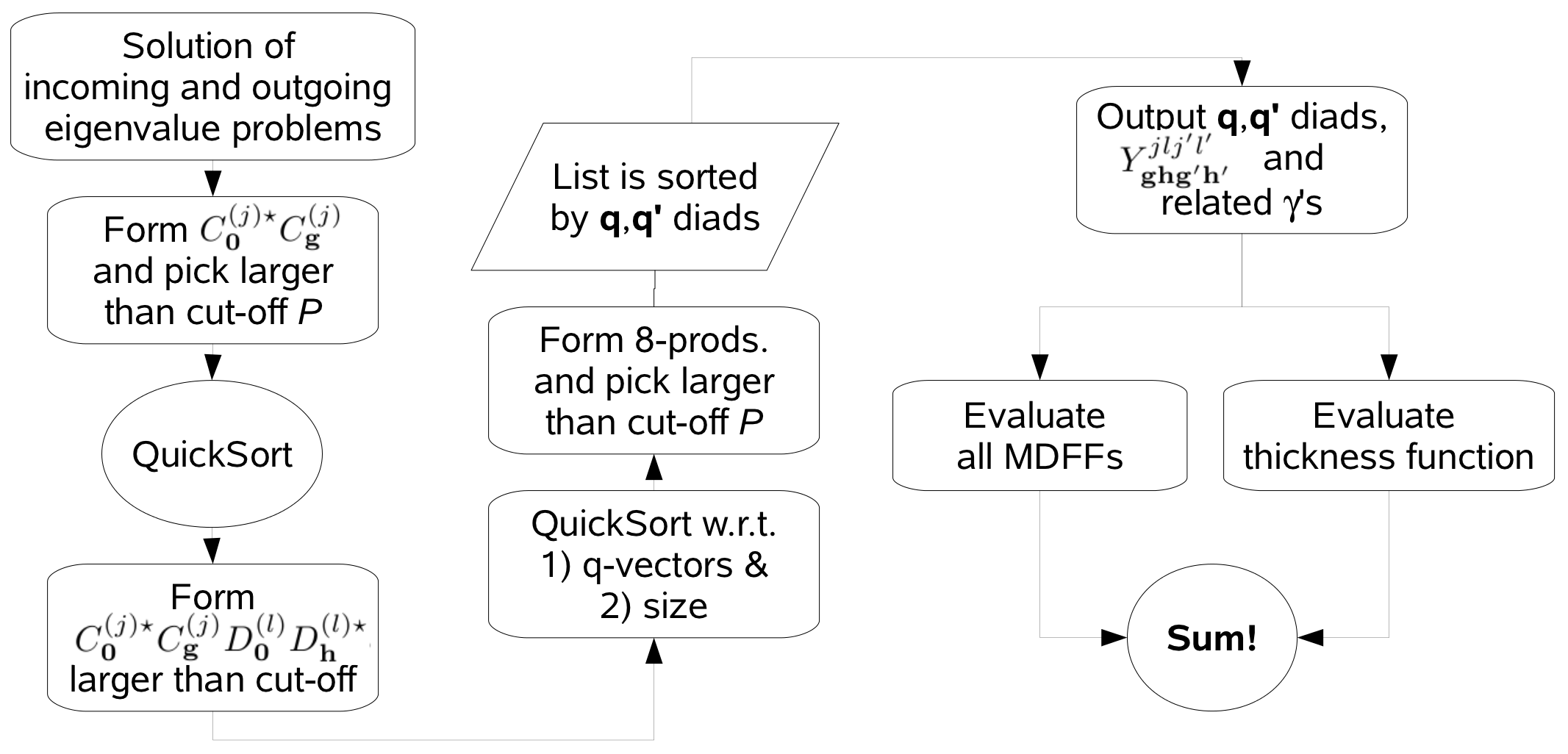}
  \caption{Flowchart of the algorithm for automatic selection of dominant terms.\label{fig:flowchart}}
\end{figure}

The new methods described below are designed to 1) avoid summation of negligible terms, 2) improve the scaling of the summation. They are built on top of the $w_\mathbf{g}$ based selection of the beams for diagonalization, assuming that we take a sufficiently large set of beams for the secular equation, typically hundreds up to few thousands. New development is in the algorithm for selection of terms to be summed from the large set of beams and Bloch waves, where no term is a priori rejected.

The algorithm is controlled by a single cut-off criterion $P_\mathrm{min}$. In the first step we create lists of diads of the Bloch coefficients, $|C_0^{(j)} C_\mathbf{g}^{(j)}| > P_\mathrm{min}$ and $|D_0^{(l)} D_\mathbf{h}^{(l)}| > P_\mathrm{min}$, for incoming and outgoing beam, respectively. Since none of $|C_\mathrm{g}^{(j)}|$ or $|D_\mathbf{h}^{(l)}|$ is larger than one, we can safely ignore all Bloch waves, for which $|C_0^{(j)}| < P_\mathrm{min}$ or $|D_0^{(l)}| < P_\mathrm{min}$. The resulting array is sorted according to decreasing magnitude using QuickSort algorithm. In total, this step has computational complexity $O(N_\mathbf{g}^2,N_2 \log N_2)$, where $N_2$ is length of the list of diads.

The next step is creation of a list of quadruple products $| C_0^{(j)} C_\mathbf{g}^{(j)} D_0^{(l)} D_\mathbf{h}^{(l)} | > P_\mathrm{min}$. This operation is $O(N_4)$, where $N_4$ is the number of the quadruples larger than $P_\mathrm{min}$, because the lists of diads were sorted by magnitude. The maximum number of failed comparisons is $O(N_2)$, where $N_2$ is the length of the longer of the two lists of diads formed in previous step.

The list of quadruples is again sorted with QuickSort algorithm, but this time first by the $\mathbf{q}$-vector given by $\mathbf{h}-\mathbf{g}$ and then by magnitude. This operation costs $O(N_4 \log N_4)$ operations. Now we have prepared data for the final step, which is an identification and output of octuple products larger than $P_\mathrm{min}$ and, simultaneously, output of the $\mathbf{q,q'}$ diads for the MDFF calculation. Thanks to the way, how we sorted the list of quadruples, we can serially process the list and output $\mathbf{q,q'}$ and corresponding octuples without actually holding the array of octuples in memory. The number of operations is $O(N_8)$, where $N_8$ is the number of octuples larger than $P_\mathrm{min}$. Maximum number of failed comparisons is given by the number of $\mathbf{q,q'}$ diads, which is well below $N_8$.

As a graphical summary of the main steps of the algorithm, a schematic flowchart diagram has been plotted in Fig.~\ref{fig:flowchart}.

Memory requirements are very favorable. Most often, the largest arrays are the $N_\mathbf{g} \times N_\mathbf{g}$ matrices used in the secular equation. The lists of quadruples rarely reach this lengths, only perhaps for extremely small $P_\mathrm{min}$ of the order below $10^{-6}$. The octuples are never held in memory, since the array is created and output serially based on the favorable sorting of the list of quadruples.

As will be shown below, $N_8$ can be orders of magnitude below the $N_\mathbf{g}^4 N_j^4$, even if naturally it has to be proportional to that. Its value strongly depends on $P_\mathrm{min}$ and it turns out that there is an approximately inverse proportionality between, i.e., $N_8 \propto P_\mathrm{min}^{-1}$. Tuning the $P_\mathrm{min}$ allows to find a suitable compromise between the speed and accuracy, which will be discussed below. Also note that we are selecting largest terms from a large set of beams, of the order of 1000, which is not feasible to treat by above-mentioned methods. Therefore it can happen that stricter selections of beams + including all cross-terms can lead to less accurate calculations, because some singular contributions of  considerable magnitude beams can be missed. Examples of such situations will be shown below.

\subsection{Automatic term selection with MDFF asymptotic\label{sec:mats}}

A modification of the algorithm is possible, if we take an advantage of the dipole-type asymptotic of the MDFFs and the $1/q^2q'^2$ denominator. For large $\mathbf{q}$-vectors the denominator suppresses the terms, therefore we can do an even more efficient rejection of the negligible terms. 

In the dipole approximation MDFF is proportional to
\begin{equation}
  S(\mathbf{q},\mathbf{q'},E) \propto \mathbf{q} \cdot \bar{N}(E) \cdot \mathbf{q'} + (\mathbf{q} \times \mathbf{q'}) \cdot \mathbf{M}(E)
\end{equation}
where $\bar{N}$ is a energy-resolved tensor dependent on density of states, local anisotropies and spin-orbit coupling and $\mathbf{M}$ is an energy-resolved vector function of local magnetic properties \cite{opmaps}. Ignoring the energy dependence, the asymptotic behavior of MDFF as a function of $\mathbf{q,q'}$ vectors is $S(\mathbf{q},\mathbf{q'}) \propto q q'$. Combining this with the denominator $1/q^2q'^2$ accompanying every MDFF, we obtain $1/qq'$ asymptotic behavior of terms.

In order to keep a dimensionless variable for the cut-off criterion, we attach to the lists of quadruples a factor $q_0/q_\mathbf{gh}$, where $\mathbf{q}_0 = \mathbf{k}_f - \mathbf{k}_i$ and $\mathbf{q}_\mathbf{gh} = \mathbf{q}_0 + \mathbf{h} - \mathbf{g}$. A small complication arises from this choice, since it is possible that the ratio $q_0/q_\mathbf{gh}$ is larger than one for some larger scattering angles or energy losses. For such eventualities we need to make sure that our lists of diads have a `reserve', i.e., the cut-off for the list of diads needs to be reduced. We have implemented a cut-off $P_\mathbf{min}/10$ for the list of diads, which is an arbitrary choice, yet it turned out to be both safe and not too costly, when compared to other list operations in the algorithm. The rest of algorithm is unchanged, only when outputting the list of selected octuples, we remove the asymptotic factor $q_0^2/q_\mathbf{gh}q_\mathbf{g'h'}$. 

In the rest of the article, we use this modified automatic term selection algorithm (MATS).

\section{Results}

In this section we compare the various methods of performing the Bloch waves summation and demonstrate some of new possibilities offered by the MATS.

\subsection{Weakly excited spots}

\begin{figure}
  \includegraphics[width=8.5cm]{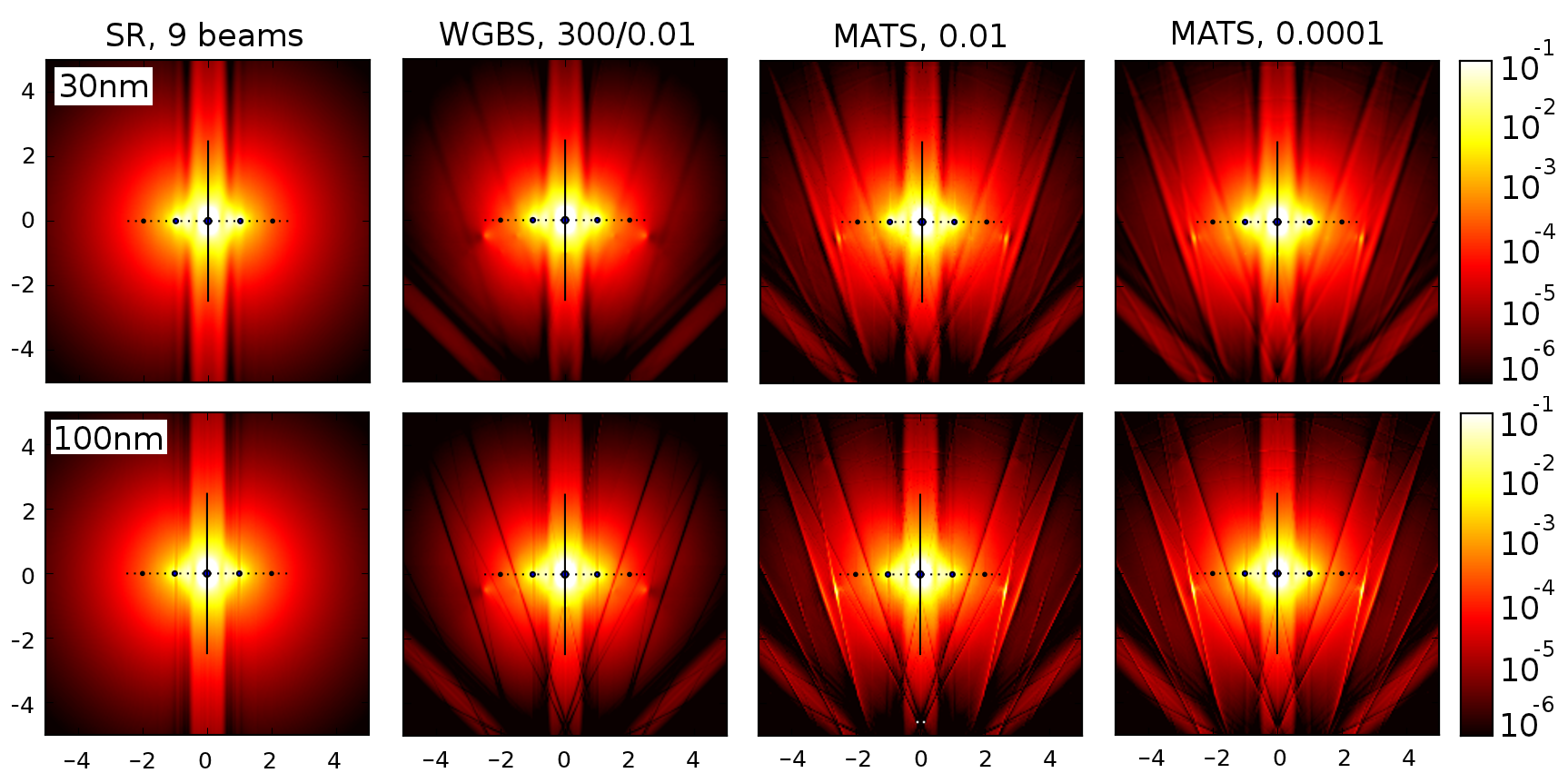}
  \caption{Diffraction pattern of iron in 3-beam orientation with $\mathbf{G}=(200)$. Top row corresponds to 30nm sample thickness, bottom row is at 100nm. From left to right various methods of calculation have been used, namely systematic row (SR) approximation, $w_\mathbf{g}$-based beam selection (WGBS) and automatic term selection with MDFF asymptotics (MATS) with two different cut-offs (see text for details). Intensities are on a logarithmic scale.\label{fig:compare}}
\end{figure}

In systematic row geometries, many simulations have been performed by only choosing beams along the systematic row. However, depending on the tilt of the beam, some of the spots outside the systematic row can be weakly excited. That can be easily missed in the systematic row approximation. Here we will show a simulation of a three-beam orientation for bcc iron with systematic row index $\mathbf{G}=(200)$. We will consider a beam tilt of approximately 5 degrees from $(001)$ zone axis orientation, which corresponds approximately to incoming beam orientation $(0,1,10)$.

Let's compare precision and timing of a systematic row approximation (SRA), $w_\mathbf{g}$-based beam selection (WGBS) and the MATS. In the SRA we included beams up to $\pm 4\mathbf{G}$, in total 9 beams and the summation was performed over all 9 resulting Bloch waves. In the WGBS, for the secular equation we set the $w_\mathbf{g}$ cut-off to 100000, which resulted in approximately 630 beams. For the summation, we used cut-off 300, which was fulfilled by 8--14 beams. For the Bloch waves we required the subspace norm to be larger than 0.01, which was fulfilled by 4--27 Bloch waves. Finally, in the MATS we tested for summation the following two cut-off criteria: $P_\mathrm{min}=0.01$ and $0.0001$.

The resulting diffraction patterns are summarized in Fig.~\ref{fig:compare}.

Regarding the computing costs, the fastest was the MATS simulation with $P_\mathrm{min}=0.01$, which finished in 43 CPU hours. In this calculation, larger part of the time was spent in diagonalization of the secular matrices of dimension 630. MATS calculation with $P_\mathrm{min}=0.0001$ finished in 98 CPU hours. The SRA calculation took considerably more time, despite being much less accurate: the 9-beam calculation finished in 561 CPU hours. Note that here we are diagonalizing very small matrices, only 9-by-9, therefore practically all the computing time is spent in summation. Finally, the WGBS calculation required 2800 CPU hours, almost 70-times more than MATS with $P_\mathrm{min}=0.01$, yet being of lower accuracy. While in our SRA calculation we always sum $9^8 = 43 \times 10^6$ terms, in MATS with $P_\mathrm{min}=0.0001$ it was on average only $70 \times 10^3$, in maximum reaching half million.

\subsection{EMCD and $m_L/m_S$ maps of iron\label{sec:conv}}

\begin{figure}
  \includegraphics[width=8.5cm]{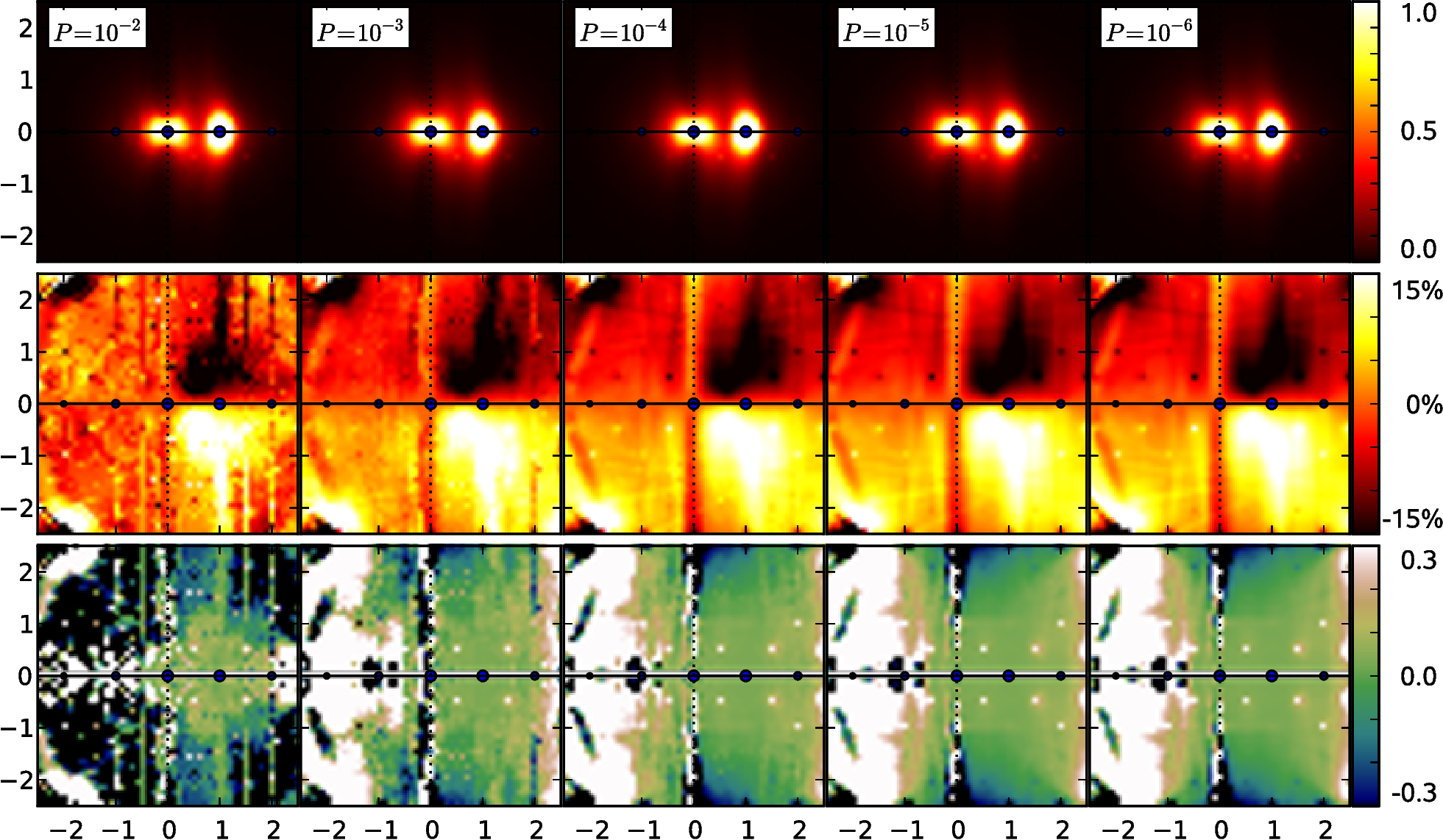}
  \caption{Diffraction patterns (top row), relative up-down difference maps (middle row) and apparent $m_L/m_S$ ratio maps (bottom row) a function of convergence parameter $P_\mathrm{min}$. Calculations were performed for 20nm layer of bcc iron in two-beam geometry with $\mathbf{G}=(200)$, at 300keV.\label{fig:conv}}
\end{figure}

Electron magnetic circular dichroism (EMCD) is a recently developed experimental technique \cite{nature}, which uses ELNES to extract the atom-specific magnetic characteristics, such as spin and orbital moments. For quantitative analyses it used sum rules \cite{oursr,lionelsr}, using which one can extract the ratio of orbital and spin moment of the studied atom \cite{hansprl}. These properties are highly sensitive functions of the edge-dependent ELNES spectra. Similarly as in experiment, also in simulations there is a high demand for precision. Below we will show the performance of MATS for the similar setup as above, two-beam case with $G=(200)$ and beam tilt of 10 degrees, showing diffraction pattern, distribution of the magnetic signal and the map of the $m_l/m_s$ ratio as a function of the cut-off variable $P_\mathrm{min}$. 

In more detail, the distribution of the magnetic signal is obtained as a difference of the diffraction pattern and its mirror image with respect to the systematic row mirror axis. In the figure, it is shown as a relative quantity, that means it is divided by the sum of the diffraction pattern and its mirror image. In other words, it is an antisymmetric part of the diffraction pattern divided by its symmetric part. Here we will not enter the discussions about the continuum signal extraction or post-edge normalization \cite{lsfollow}, since our main focus is the convergence of the Bloch waves calculation.

The maps of the $m_L/m_S$ ratio are evaluated pixel-by-pixel from the magnetic signals $M_{L_3}$ and $M_{L_2}$ integrated over $L_3$ and $L_2$ edge regions, respectively, by the following formula \cite{oursr,lionelsr}
\begin{equation}
  \frac{m_L}{m_S} = \frac{2}{3} \frac{M_{L_3}+M_{L_2}}{M_{L_3}-2M_{L_2}}
\end{equation}
Intuitively, this should be a constant function throughout the diffraction plane. However, in fact, large variations occur due to asymmetries discussed in detail in \cite{2bcasymm}. The map of the $m_L/m_S$ ratio is a highly sensitive function of the scattering cross-section and is an excellent test of the convergence properties of the newly developed method.

We have used the same $w_\mathbf{g}$ cut-off as in previous sub-section, but we have varied the $P_\mathrm{min}$ from $10^{-2}$ down to $10^{-6}$ and recorded some statistic information about the number of terms included in the summation, Table~\ref{tab:conv}. Note that, as anticipated, the number of summed terms is inversely proportional to $P_\mathrm{min}$. Importantly, even at the most accurate calculation, the number of summed terms stays many orders of magnitude below $700^8$, demonstrating the high efficiency of the selection of terms.

\begin{table}
  \caption{Average lengths of double $\langle N_2 \rangle$, quadruple $\langle N_4 \rangle$ and octuple $\langle N_8 \rangle$ product lists, average number of momentum transfer diads per energy step $\langle N_\mathbf{qq'} \rangle$ and computing times (total and per-pixel average) for maps in Fig.~\ref{fig:conv} as a function of convergence parameter $P_\mathrm{min}$. Times refer to a single Intel Pentium 4 Xeon processor at 2.5GHz.\label{tab:conv}}
  \begin{tabular}{lrrrrrr}
    \hline \hline
 $P_\mathrm{min}$ & \multicolumn{1}{c}{$\langle N_2 \rangle$} & \multicolumn{1}{c}{$\langle N_4 \rangle$} & \multicolumn{1}{c}{$\langle N_8 \rangle$} & \multicolumn{1}{c}{$\langle N_\mathbf{qq'} \rangle$} & \multicolumn{1}{c}{$t_\mathrm{total}$} & \multicolumn{1}{c}{$\langle t_1 \rangle$} \\
    \hline
      $10^{-2}$   &   230 &     72 &     330 &    12 &   2h 16min  &  3.1s \\
      $10^{-3}$   &  1355 &    750 &    4710 &   110 &   3h 18min  &  4.6s \\
      $10^{-4}$   &  5545 &   5865 &   56830 &   705 &  11h 13min  & 15.5s \\
      $10^{-5}$   & 23060 &  39760 &  570250 &  3600 &  83h 35min  & 115s  \\
      $10^{-6}$   & 73591 & 259739 & 5398841 & 16860 & 628h 14min  & 870s  \\
    \hline \hline
  \end{tabular}
\end{table}

Note that the diffaction pattern appears to be reasonably converged already for $P_\mathrm{min}=10^{-2}$. An attentive reader might spot slightly fuzziness, but nevertheless, the differences between all five diffraction patterns are visually negligible. The relative difference map requires better convergence, at least $P_\mathrm{min}=10^{-3}$, or better $P_\mathrm{min}=10^{-4}$. Note the numerical noise at $P_\mathrm{min}=10^{-3}$, particularly along the vertical line going through the transmitted beam. Although a slight fuzziness remains at $P_\mathrm{min}=10^{-3}$, but the accuracy is already satisfactory. The most sensitive quantity presented here is the apparent $m_L/m_S$ ratio. Since the orbital momentum in iron is small, the nominator in the sum rule expression is a difference of two (typically small) differences of spectra, integrated over $L_2$ or $L_3$ edge, respectivelly. It requires high accuracy to obtain well converged maps. Calculation with $P_\mathrm{min}=10^{-4}$ seems to produce reasonably converged results. The only visible difference at $P_\mathrm{min}=10^{-5}$, when compared to $P_\mathrm{min}=10^{-4}$, appears around $(-1,\pm 1)G$ and $(2,\pm 2)G$ positions. The results of calculation with $P_\mathrm{min}=10^{-6}$ are visually indistinguishable from $P_\mathrm{min}=10^{-5}$.

Looking back into the Table~\ref{tab:conv}, we see that for most of our purposes $P_\mathrm{min}=10^{-4}$ provides highly converged results at average costs 3-times lower than the simplest testing calculation discussed in Sec.~\ref{sec:wgbs}. Nevertheless, this criterion can depend on the studied crystal structure and orientation and always should be tested, when performing simulations of new systems.



\subsection{Tilting crystal from the zone axis orientation}

High efficiency of the MATS method allows us to approach experimental geometries, which require a large number of beams for a converged simulation. Particularly, the zone axis orientation is highly computationally intensive. Here we will follow the development of the diffraction pattern, when tilting from an exact zone axis towards the 3-beam case with $\mathbf{G}=(200)$ in small steps of the tilt. Extremely rich patterns are observed, particularly when looking at the distribution of the magnetic signal, with formation of a dense network of Kikuchi bands and lines.

The $w_\mathbf{g}$ cut-off was the same as in previous sections, the $P_\mathrm{min}$ was set to $10^{-4}$, leading to well converged diffraction patterns and maps of the magnetic signal, as demonstrated in Sec.~\ref{sec:conv}. We chose to demonstrate the results at a sample thickness 50nm, because at this thickness the Kikuchi patterns appear sharp enough, yet not so sharp that we would observe aliasing artefacts due to discrete grid of pixels.

The incoming beam is gradually tilted from an exact zone axis $(001)$ direction towards the 3-beam orientation via $(0,1,40) \simeq 1.43$ degrees, $(0,1,20) \simeq 2.86$ degrees, $(0,1,10) \simeq 5.71$ degrees and $(016) \simeq 9.46$ degrees tilt. At the endpoints we also did calculations with manual selection of beams. For the exact zone axis case we included kinematically allowed $\mathbf{g}$-vectors from the zero-order Laue zone (ZOLZ) with $hkl$ indices less than 4 (in total 25 $\mathbf{g}$-vectors) and for the final 3-beam orientation we included beams of up to $\pm5\mathbf{G}$ (in total 11 $\mathbf{g}$-vectors).

\begin{figure}
  \includegraphics[width=8.5cm]{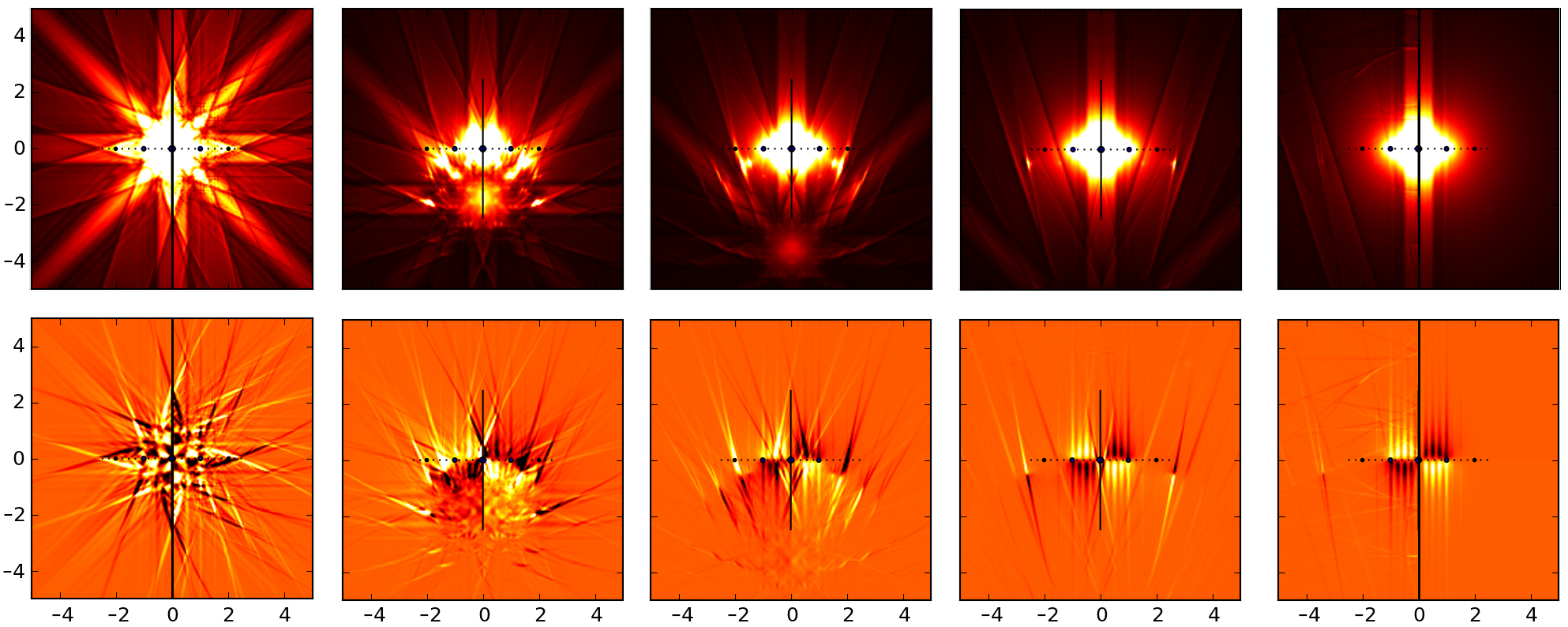}
  \caption{Evolution of the diffraction pattern and of the map of magnetic signal when tilting the incoming beam from $(001)$ zone-axis orientation to 3-beam orientation $(016)$. Left halfplane of zone axis orientation is calculated with 25-beams, right half-plane of 3-beam orientation is calculated with 11 beams on the systematic row (see text for details). Note the differences from MATS calculations.\label{fig:zatilt}}
\end{figure}

The results are summarized in Fig.~\ref{fig:zatilt}. At the zone axis orientation we see a rich pattern, consisting of Bragg spots and multitude of Kikuchi bands and lines. The Kikuchi pattern is even better visible on the map of the magnetic signal. Number of thin lines with varying intensity and sign form a rich symmetric structure. Now let's compare them to a calculation with 25 $\mathbf{g}$-vectors from ZOLZ. The diffraction pattern seems to display the same structure, except for higher intensity. A cautious eye would spot some differences in relative intensity around $\{ 220 \}$ spots. More differences can be seen in the maps of the magnetic signal. Particularly at larger scattering angles, the pattern of Kikuchi lines is more rich in the MATS description.

As we tilt the beam, the pattern of beams is deforming, following the movement of the zone axis spot down. At beam tilt $1.43$ degrees, corresponding to approximately 25~mrad, the zone axis spot moved down by a bit less than $2G$. The two-fold Bragg angle corresponding to $\mathbf{G}=(200)$ in bcc iron is 13.8~mrad, therefore the $2G \approx 27.6$~mrad, which agrees with the position of the zone axis spot in the map. The dominant Kikuchi bands are passing around this spot. In the map of the magnetic signal a noise-like signal forms around the zone axis center, and at the same time, there evolves a different dominant sign in the four quadrants of the diffraction plane. Both these trends continue with the tilt $2.86$ degrees. The zone axis spot has weaker intensity, moves further down, to around $3.5G$ under the transmitted beam. At this angle and thickness, we see relatively strongly excited beams for $h=-3,-1,1,3$ and $k=-1$, i.e., spots under the systematic row of reflections. At $5.7$ degrees tilt the zone axis spot is outside the vertical range of our maps, we see a clean 3-beam pattern, with some intensities at $(\bar{1}\bar{1}0)$ and $(1\bar{1}0)$. Maps of the magnetic signal are considerably simpler, having a dominant sign in each quadrant. The four vertical lines represent Kikuchi lines. Skew lines at larger scattering angles show unexpectedly high magnetic signal, which might be a finding of potential practical importance.

At the final step, almost $10$ degrees tilt the patterns complete the trends: three strongly excited spots along the systematic row, and three dominant Kikuchi bands. Magnetic signal showing the four vertical lines with signs corresponding to their quadrants. Interesting is a comparison to a systematic row approximation, where we used only 11 $\mathbf{g}$-vectors -- multiples of the $\mathbf{G}=(200)$. These patterns completely miss the skew Kikuchi bands and in the maps of the magnetic signal there is a very reduced pattern of lines.

\subsection{Comparison between MATS and ICSC results}

A similar but alternative computer program which calculates inner-shell ionization and backscattering cross sections for fast electrons incident on a crystal is available, known as the ICSC code, developed by Oxley and Allen \cite{icsc}. The program calculates the inelastic scattering coefficients for inner-shell ionization, pertinent to EELS and energy dispersive X-ray (EDX) analysis, using parameterizations of the atomic inelastic scattering factors. The program treats the dynamical scattering in a very similar way to MATS, but simplifies the calculations of dynamical form factors, using the approximation where the integration over all the final states of the scattered electron is replaced by an analytic expression, which unfortunately does not cover MDFF calculation. This means that the ICSC code only allows the EELS detector geometry concentric with the incident beam direction. The ICSC code has been intensively tested for EDX results of atom location by channeling enhanced microanalysis (ALCHEMI), but little was published in testing the ICSC program for experimental EELS. It is thus a good chance for comparing between the two methods.

As a benchmark we selected a cubic SiC crystal and the orientation dependent Si K-edge cross sections were compared. We have done two sets of calculations: 1) thickness dependence in a beam-rocking experiment in a systematic row orientation, and, 2) 2-dimensional beam-rocking in a (001) zone-axis orientation. In both cases, we selected parallel illumination and convergence angle of 10~mrad. The acceleration voltage was set to 300~keV. ICSC uses Doyle-Turner scattering factors \cite{doyle}, therefore for the sake of consistency, we used these also in calculations with the new code.

\begin{figure}
  \includegraphics[width=8.5cm]{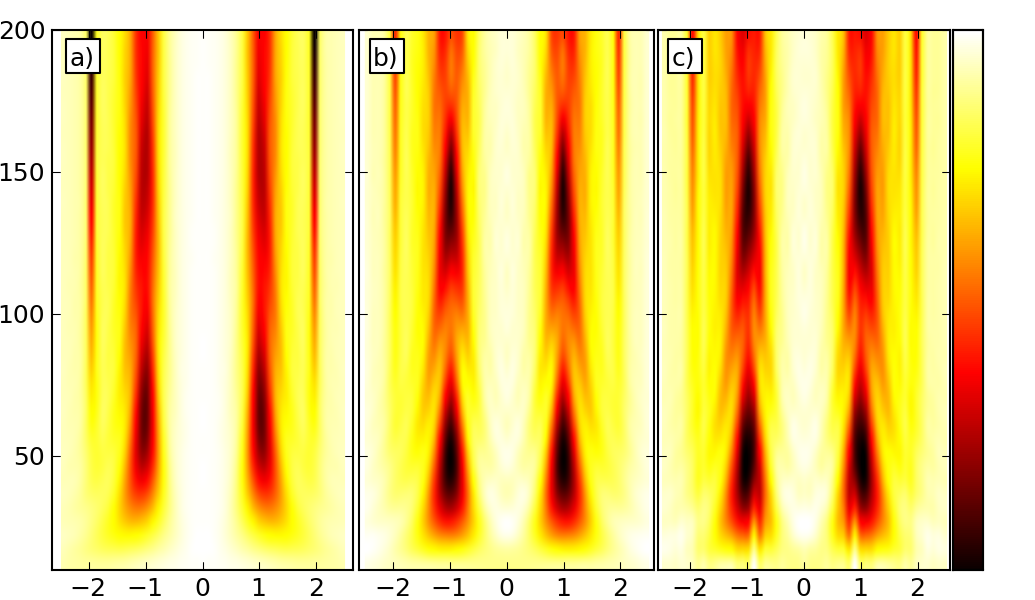}
  \caption{Comparison of ICSC and MATS. Thickness profiles of K-edge of Si in SiC in systematic row orientation. Horizontal axis corresponds to a beam tilt in multiples of $\mathbf{G}=(220)$ and the vertical axis is thickness in nm. a) ICSC calculation, b) calculation with the new code, adopting the same parameters as ICSC, c) MATS calculation.\label{fig:icsc_sr}}
\end{figure}

In the systematic row calculation, we assumed a beam tilt of approximately 10 degrees towards the $G_{(220)}$ systematic row conditions, i.e., the zone axis was $(\bar{1}18)$. For ICSC we used as set of 85 beams $(hkl) \perp (\bar{1}18)$. With the new code we did two sets of calculations - one with the fixed set of 9 beams from the systematic row (SR; up to $\pm 4 G_{(220)}$) and another in the MATS approach with convergence parameter $10^{-4}$. Results are summarized in Fig.~\ref{fig:icsc_sr}. The results show good qualitative agreement, yet there are clear differences in details, even when comparing the ICSC calculation to a SR calculation. That indicates that approximations introduced in ICSC partially neglect some details of the dynamical diffraction process. Full MATS calculation shows expectedly even more of a fine structure, especially at larger thicknesses. Detailed comparison to experiment would be needed to verify the fine features observed in MATS calculation.

\begin{figure}
  \includegraphics[width=8.5cm]{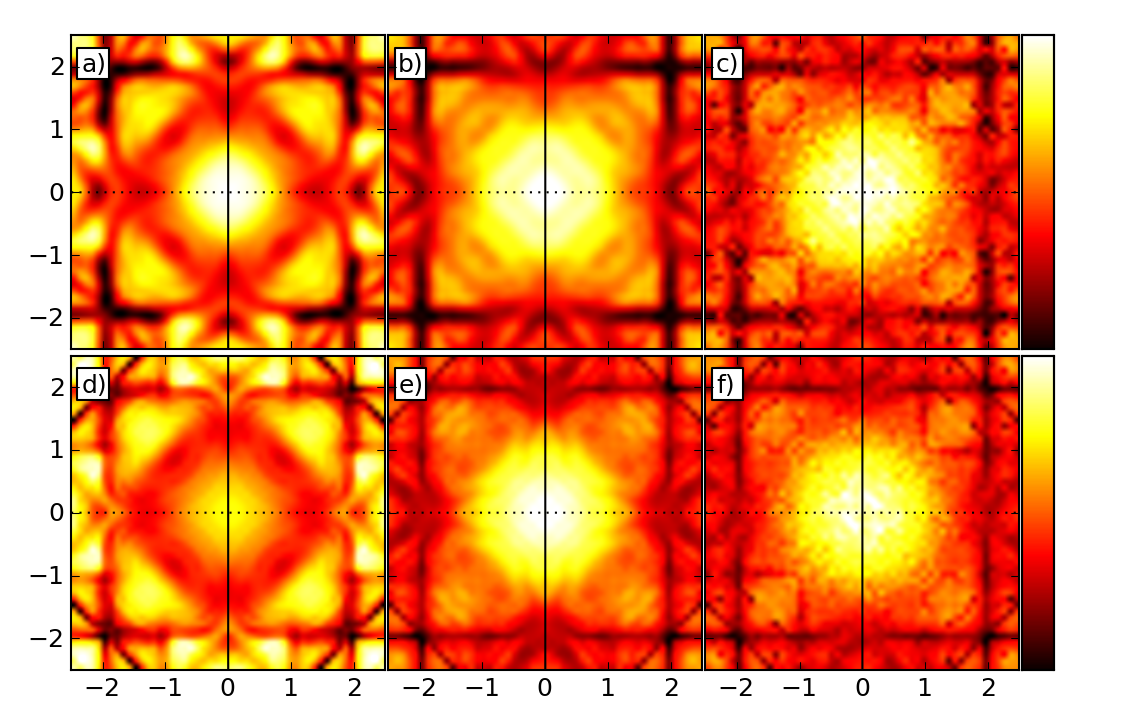}
  \caption{Comparison of ICSC and MATS. Intensity of K-edge of Si in SiC in $(001)$ zone-axis orientation as a function of beam rocking. Axes corresponds to a beam tilt in multiples of $\mathbf{G}=(200)$. a) ICSC calculation at 100~nm, b) calculation with the new code, adopting the same parameters as ICSC, c) MATS calculation at 100~nm, d)-f) the same as a)-c), but at 200~nm.\label{fig:icsc_za}}
\end{figure}

The dependence of the double-differential scattering cross-section on beam rocking from the $(001)$ zone axis orientation is shown in Fig.~\ref{fig:icsc_za}. The ICSC used results are shown in the left column. A set of 197 beams was used, all from the zero-order Laue zone. The same ones were used for comparison with the new code (shown in the middle column). Finally, results obtained by the MATS method are shown in the right column. General qualitative features are in good agreement between all three computational approaches, nevertheless, as in the systematic-row orientation, also here we observe notable differences. Of particular interest is a soft breaking of the four-fold symmetry in the case of MATS calculations. It is easier to spot at 100~nm, where the central intensity maximum has a slighly prolate shape along the main diagonal of the pattern. This can be explained by presence of beams from higher-order Laue zones in the secular equation. They introduce $(hkl)$ beams with $l \ne 0$ and due to the curvature of the Ewald sphere, $(hkl)$ and $(hk\bar{l})$ have different excitation errors and thus their corresponding Bloch coefficients differ. In combination with the four-fold screw axis of this structure and a lack of inversion symmetry, the resulting beam-rocking pattern shows deviation from four-fold symmetry.

\section{Conclusions}

We have developed a new method for accurate summation over Bloch waves and their plane wave components (beams) named modified automatic term selection (MATS). The complexity of MATS scales inversely proportionally with the cut-off for the term sizes. It allows highly accurate calculations at much lower computational costs compared to previous methods. We have demonstrated advantages of the method on capturing the intensity of the weakly excited spots outside the systematic row. The convergence properties of MATS were studied on a two-beam case simulation with focus on faint effects observed in EMCD experiments. A rich pattern of Kikuchi bands and lines was presented in a simulation of a tilting of the bcc crystal from the zone axis orientation to a three-beam orientation. We have also compared the new method to ICSC code and both display qualitatively the same results. In more detail, MATS calculation seems to provide more rich structures, most likely due to larger number of beams included in the secular equation.

\section{Acknowledgements}

J.R.\ acknowledges the support of Swedish Research Council and STINT. Simulations were performed on computer cluster \textsc{david} of Czech Academy of Sciences. This work was in part supported by the Grant-in-Aid for Scientific Research from MEXT, Japan.

\end{document}